\documentclass[preprint,amsmath,amssymb]{revtex4}

\usepackage{graphicx,bm}

\begin{document}

{\bf Comment on the ``Novel Isotope Effects Observed in Polarization
Echo Experiments in Glasses''}

In a recent letter \cite{NFHE}, Nagel at el. have conclusively
demonstrated that the remarkable, unexpected effect of magnetic field
on the low temperature anomalies in some amorphous substances arises
from the coupling of the nuclear quadrupole moments (QM) to the
residual anharmonic degrees of freedom commonly referred to as the
two-level systems (TLS).  This was achieved by comparing the
dependences of the spontaneous echo decay rate on the magnetic field
strength in regular and fully deuterated amorphous glycerol.
Quadrupole splitting introduces an additional source of relaxation,
thus reducing the echo amplitude by about 50\% in the deuterated
sample. A magnetic field of sufficient strength restores the signal
(see below).  But even in regular glycerol some magnetic field effects
were observed, albeit weaker in magnitude: about a 3\% echo reduction
is observed. Based on the natural abundance of the $^{17}$O and $^2$H
isotopes (possessing QM) and assuming that a {\em single} QM per TLS
will account for the effect, Nagel at el. conclude that ``on average
more than ten glycerol molecules are involved in an individual
tunneling process'', subject to the potential presence of other QM
carrying impurities. Far from a surprise, this conclusion is expected
on theoretical grounds.

We first point out that the TLS are expected to be the low energy
limit of multilevel excitations remaining in the liquid after it is
quenched below $T_g$, its glass transition temperature \cite{LW}. Such
excitations have been predicted to involve about 200 independent
molecular units, or ``beads'', undergoing strongly anharmonic motions
between two alternative local structural states. Tunneling occurs
stagewise by moving a ``domain wall'', separating those alternative
structural states, while individual atoms experience modest
displacements of the order the Lindemann ratio $\rho_L \simeq 0.1$ of
the bead's size.  At somewhat higher temperatures the domain wall
vibrations become thermally activated and give rise to the Boson Peak
phenomena \cite{LW_BP}.

The results of Ref.\cite{NFHE} are quite consistent with this theory.
The size of the independently moving units in a liquid can be deduced
from the entropy loss upon crystallization, or by detailed fits of the
$T$-dependence of the viscosity above $T_g$ \cite{LW_soft}. Using
glycerol's fusion entropy per molecule $\sim$3.5 $k_B$ indicates that
each molecule of C$_3$H$_8$O$_3$ has about 3.5 beads, implying roughly
one bead per C-O unit. This is in good accord with chemical
intuition. Since the maximum QM induced echo reduction by magnetic
field is only $\sim$50\%, the estimate of the size of a rearranging
region of Nagel at el. should actually be more than 20 glycerol
molecules per TLS rather than 10. This lower bound gives about $N_b
\sim 70$ beads per tunneling center (TC). We next present two pieces
of {\em direct} evidence of the mesoscopic character of a TC.

The echo ``beating'' frequency from Fig.3 of Ref.\cite{NFHE}, equal to
$\sim$0.14 MHz, was identified by Nagel et al. as ``reflecting
directly the mean quadrupole splitting'' on a proton site $\delta_Q$,
because it is close to the experimentally measured
$\delta_Q$(C-$^2$H)$/2 \pi= 0.158$ MHz and $\delta_Q$(O-$^2$H)$/2 \pi
= 0.124$ MHz. On the contrary, the beating frequency should reflect
the {\em total} QM induced splitting $\Delta_Q$ of the whole TC, as in
the analogous problem of electron spin rotation in radicals with many
hyperfine coupled nuclei \cite{KlausPeter}. Since every bond rotates
only about $\rho_L \sim 0.1$ radians during a transition, individual
QM contributions to the nutation, $\rho_L \delta_Q$ are small and
random; however in deuterated glycerol the number of deuteriums in a
TC is large: $N_D \simeq (N_b/3.5) \cdot 8 \simeq 460$, using the
theoretical $N_b=200$. Since the bond rotations are randomly oriented,
one gets $\Delta_Q \sim \delta_Q \rho_L \sqrt{N_D}$. The resulting
beating frequency is $(\Delta_Q/2)/2\pi \sim 0.14$ MHz!  Conversely,
associating the beating frequency $0.14$ MHz with an {\em individual}
quadrupole splitting would imply a very large angle bond rotation
during a tunneling transition.

We stress the difference in echo reduction between normal and
deuterated glycerol is not only an effect on the magnitude of the
signal. According to Figs. 1 and 2 of Ref.\cite{NFHE}, a much weaker
field suffices to restore the echo amplitude in the regular sample, a
factor of 10 in comparison with the deuterated sample.  The echo
reduction can be understood as arising from interfering transitions in
a large Hilbert space, which is a direct product of the structural
transition and the many precessing nuclear spins. The structural
states are mixed with the nuclear ones because the electric field
gradients are oriented differently in the two structural states.
Having {\em one} QM per TC results in an effective quadrupole
splitting of the tunneling center less, by a factor of $N_D^{-1/2}$,
than in the deuterated sample because the total ``band width'' of $N$
random off-diagonal couplings, such as provided by the QM's, scales as
$N^{1/2}$ \cite{Mehta}, also consistent with the geometric argument
above.  The Zeeman splitting on {\em each} nucleus must exceed this
band width by some fixed factor in order to fully energetically
separate the distinct nuclear precessions. Assuming $^{17}$O and $^2$H
impurities in regular glycerol contribute equally, the ratio of
characteristic fields gives $N_D \sim 3 \cdot 10^2$.

That tunneling centers should involve moving only very few atoms is
actually rather unlikely on energetic grounds: Suppose for a moment,
in an extreme fashion, that the TC were indeed very small, i.e. one or
two molecular units across. In order to have a {\em resonant}
transition with a reasonable tunneling amplitude, so that it is active
at low $T$, there must be available a ``free volume'' comparable to
the bead's volume itself. Since the Lindemann ratio is $\sim$0.1 at
crystallization or vitrification alike, the typical specific volume
fluctuations at $T_g$ are about $\sqrt{3} \cdot 0.1$, i.e. 20\% or
less. Density fluctuations of order 100\% have at best
$\exp[-(100/20)^2/2] \sim 10^{-6}$ probability! Indeed, point-like
defects are rare \cite{EastwoodW} - at least as rare in glasses as in
crystals. The single bead defect density, as estimated above, would be
much less than the TLS density needed to explain experiment.  Such
simplistic, ``free-volume'' reasoning does not apply to structural
transitions encompassing any larger regions, of course, because more
complicated motions are allowed thereby that are accompanied by
smaller density changes (see details in \cite{XW,LW_soft}).

\vspace{1mm}

V. Lubchenko 

\hspace{5mm} \begin{tabular}{l}
Department of Chemistry, \\
Massachusetts Institute of Technology \\
Cambridge, MA 02139
\end{tabular} \\

P. G. Wolynes 

\hspace{5mm} \begin{tabular}{l}
Departments of Chemistry and Biochemistry, and Physics \\
University of California San Diego \\
La Jolla, CA 92093-0371
\end{tabular} \\

\end{document}